\documentclass[11pt,prb,reprint,superscriptaddress,citeautoscript]{revtex4-1}
\usepackage[caption=false]{subfig}
\usepackage[version=4]{mhchem}
\usepackage{graphicx,color}
\usepackage[usenames,dvipsnames,svgnaes,table]{xcolor}
\usepackage{bm}
\usepackage{amsmath}
\usepackage{amssymb}
\usepackage{mathtools}
\usepackage{siunitx}
\usepackage[super]{nth}
\usepackage{upgreek}
\usepackage{multirow}
\usepackage{array}
\usepackage{color}
\usepackage{longtable}
\usepackage{xfrac}
\usepackage{hyperref}
\hypersetup{colorlinks=true,
            linkcolor=red,
            citecolor=blue,
            urlcolor=orange}
\usepackage{graphicx, type1cm, lettrine}
\newcolumntype{K}[1]{>{\centering\arraybackslash}p{#1}}
\setcitestyle{numbers,super}

\usepackage{ amssymb }
\usepackage{ amsmath }
\usepackage{ amsfonts}
\usepackage{ dsfont }
\usepackage{ amsthm}
\usepackage{algpseudocode, algorithm}
\usepackage{bbm}
\usepackage{ enumerate}
\usepackage{bm}

\newcommand{\N}{\mathcal{N}}
\newcommand{\GP}{\mathcal{GP}}

\newcommand{\sai}{$\Upsigma_{\text{AI}}$}

\newcommand{\chiai}{$\text{X}_{\text{AI}}$}

\AtBeginDocument{\usepackage{booktabs}}

\begin{document}
\title{Autonomous synthesis of metastable materials}

\author{Sebastian Ament}
\affiliation
{Department of Computer Science, Cornell University, Ithaca, NY 14853, United States}
\affiliation
{These authors contributed equally to this work}
\author{Maximilian Amsler}
\email{amsler.max@gmail.com}
\affiliation{Department of Materials Science and Engineering, Cornell University, Ithaca, NY 14853, United States}
\affiliation
{Department of Chemistry and Biochemistry, University of Bern, Freiestrasse 3, CH-3012 Bern, Switzerland}
\affiliation
{These authors contributed equally to this work}
\author{Duncan R. Sutherland}
\affiliation{Department of Materials Science and Engineering, Cornell University, Ithaca, NY 14853, United States}

\author{Ming-Chiang Chang}
\affiliation
{Department of Materials Science and Engineering, Cornell University, Ithaca, NY 14853, United States}

\author{Dan Guevarra}
\affiliation
{Joint Center for Artificial Photosynthesis, California Institute of Technology, Pasadena, CA 91125}

\author{Aine B. Connolly}
\affiliation{Department of Materials Science and Engineering, Cornell University, Ithaca, NY 14853, United States}

\author{John M. Gregoire}
\affiliation
{Joint Center for Artificial Photosynthesis, California Institute of Technology, Pasadena, CA 91125}

\author{Michael O. Thompson}
\affiliation{Department of Materials Science and Engineering, Cornell University, Ithaca, NY 14853, United States}

 \author{Carla P. Gomes}
\email{gomes@cs.cornell.edu}
\affiliation
{Department of Computer Science, Cornell University, Ithaca, NY 14853, United States}

\author{R. Bruce van Dover}
\affiliation{Department of Materials Science and Engineering, Cornell University, Ithaca, NY 14853, United States}

\date{\today}

\begin{abstract}
Autonomous experimentation enabled by artificial intelligence (AI) offers a new paradigm for accelerating scientific discovery. Non-equilibrium materials synthesis is emblematic of complex, resource-intensive experimentation whose acceleration would be a watershed for materials discovery and development. The mapping of non-equilibrium synthesis phase diagrams has recently been accelerated via high throughput experimentation but still limits materials research because the parameter space is too vast to be exhaustively explored. We demonstrate accelerated synthesis and exploration of metastable materials through hierarchical autonomous experimentation governed by the Scientific Autonomous Reasoning Agent (SARA). SARA integrates robotic materials synthesis and characterization along with a hierarchy of AI methods that efficiently reveal the structure of processing phase diagrams. SARA designs lateral gradient laser spike annealing (lg-LSA) experiments for parallel materials synthesis and employs optical spectroscopy to rapidly identify phase transitions. Efficient exploration of the multi-dimensional parameter space is achieved with nested active learning (AL) cycles built upon advanced machine learning models that incorporate the underlying physics of the experiments as well as end-to-end uncertainty quantification. With this, and the coordination of AL at multiple scales, SARA embodies AI harnessing of complex scientific tasks. We demonstrate its performance by autonomously mapping synthesis phase boundaries for the \ce{Bi2O3} system, leading to orders-of-magnitude acceleration in the establishment of a synthesis phase diagram that includes conditions for kinetically stabilizing  $\delta$-\ce{Bi2O3} at room temperature, a critical development for electrochemical technologies such as solid oxide fuel cells.
\end{abstract}

\maketitle

\section{Introduction}
Artificial intelligence (AI) holds great promise for revolutionizing scientific fields as varied as biology~\cite{senior2020improved}, chemistry~\cite{huang_quantum_2020}, physics~\cite{carleo_solving_2017}, and economics \cite{jean2016poverty}.
Much of AI's impressive recent successes have been in data-rich applications:
AlphaFold \cite{senior2020improved}, for example, uses a library of tens of thousands of existing protein data to build a highly successful model for protein folding.
In fields that lack comparably vast data, however, 
a great opportunity lies in guiding the exploratory process 
itself to minimize the number of experiments that are required to achieve insights, i.e, active learning~\cite{settles_active_2009} (AL), and thereby accelerate the pace of scientific discovery \cite{gil2014amplify}.

Such AI-guided efforts
have shown great promise in the design of quantum experiments \cite{melnikov1221}, drug development \cite{murphy2011active},
wind turbine control \cite{kolter2012design},
and 
are of particular importance in materials research aimed at designing and optimizing functional materials that lie at the core of technological advances.
High-throughput (HT) experimental synthesis and characterization of materials systems through thin-film deposition of inorganic composition spreads~\cite{dover_codeposited_2004}, so-called ``libraries'', present a promising avenue to rapidly explore a vast chemical, structural, and property space~\cite{hattrick-simpers_perspective_2016}. 
These methods have been well established for comprehensive synthesis of composition spaces with 2-4 components, where the resulting 10's to 1000's of materials can be evaluated via automated characterization. 
While this approach has been quite effective for identification of materials with desired properties, the opportunity for broader materials exploration to enable new technologies is highlighted by the limited exploration of synthesis conditions to-date. 
The portion of the materials search space that has been explored is vanishingly small when considering the dynamic range of thermal processing conditions, which are inherent to  processing-composition-structure-property (PCSP) relationships. 
The breadth of relevant thermal processing conditions makes exhaustive sampling untenable. 
Therefore, AL is critical in reducing the number of experiments to a more tractable scale~\cite{reyes_machine_2019, sstein_progress_2019, tabor_accelerating_2018}.

Recently, HT experimentation and AL techniques have been combined in a closed-loop fashion, where an AI instance iteratively proposes a sequence of experiments to explore and discover new materials.
These efforts include identifying phase-change materials via Bayesian AL~\cite{kusne_on-the-fly_2020}, the discovery of NiTi-based shape memory alloys with low thermal hysteresis~\cite{xue_accelerated_2016}, the synthesis of \ce{BaTiO3}-based piezoelectrics with the large electrostrain~\cite{yuan_accelerated_2018}, 
the selective growth of carbon nano-tubes~\cite{nikolaev_discovery_2014,nikolaev_autonomy_2016},  the search for perovskite-type materials for photovoltaic applications~\cite{sun_accelerated_2019}
and inorganic quantum dots~\cite{epps_artificial_2020}, 
maximizing hole mobility of organic solar cells~\cite{macleod_self-driving_2020},
and
accelerating thoughness-optimization in additive manufacturing~\cite{gongora_bayesian_2020}.

Despite this progress, the current state of the art exhibits considerable limitations.
Chiefly, most attempts at closed-loop cycles still rely heavily on human intervention, preventing them from reaching true autonomy in materials discovery.
Further, although AL guidance has recently been deployed to great effect for discovery of optical phase-change thin film materials,~\cite{kusne_on-the-fly_2020} the search space was limited to pre-synthesized compositions using a single processing condition. 
The time and temperature scales relevant to non-equilibrium thermal processing of solid-state inorganic materials pose substantial problems for incorporating synthesis in an autonomous loop, although the utility of spanning synthesis and characterization in an AL framework has been clearly demonstrated by several recent autonomous workflows for chemical synthesis
~\cite{porwol_autonomous_2020,
li_autonomous_2020,
macleod_self-driving_2020,
li_robot-accelerated_2020}.
The complexity and the degrees of freedom of the PCSP space are particularly challenging to incorporate in autonomous experimentation when considering metastable materials that form far from equilibrium at different, often unpredictable processing conditions~\cite{alberi_2019_2018, saksena_metastable_2018}.
Importantly, commonly deployed off-the-shelf AL 
models are often not sufficient for achieving highly efficient learning and
are frequently outperformed by random search with twice the number of samples~\cite{li2017hyperband},
a problem that is exacerbated with increased dimensionality of the search space~\cite{ahmed2016we}.

Expert human scientists navigate complex search spaces by incorporating their prior knowledge, such as physics-based models that underlie the acquired data. 
Incorporating such knowledge in AL often requires the development of new AI methods. 
Finally, exploration via AL critically relies on uncertainty quantification in the not-yet-sampled regions of parameter space, which for complex experimental workflows requires error propagation. 
Arguably the most immediate obstacle to accelerating experimental exploration via AL lies in the dual challenges of developing noise models for each type of experiment and integrating them into a computational framework for end-to-end uncertainty quantification. 
In aggregate, these challenges motivate the establishment of a framework that integrates AI methods at multiple scales to perform scientifically meaningful interpretation, modeling, and uncertainty quantification of multiple streams of incoming data.

Our vision of the Scientific Autonomous Reasoning Agent (SARA)~\cite{sara2020vision} is to develop a fully autonomous HT materials discovery and exploration framework by integrating robotic HT materials synthesis~\cite{li_robot-accelerated_2020} with AI instances
to accelerate both materials synthesis and analysis. 
In particular, SARA aims to automate the representation, characterization, planning, optimization, and learning of materials knowledge in a fully integrated manner. To achieve this goal, we envision the deployment of agents, which individually specialize on specific subtasks, but closely interact with each other to accelerate the discovery efforts.
These agents include, but are not limited to, synthesis and probing robotics to conduct experiments 
and highly optimized, physics-based AI models that evaluate currently available data with their associated uncertainties and that drive AI-guided discovery.

In this work, we take strides towards realizing this vision and present a fully integrated, autonomous framework that iteratively maps out the synthesis phase boundaries of metastable compounds in a closed-loop fashion.
{To this end, we incorporate a system of nested~\cite{balachandran_experimental_2018} 
cycles harnessed by SARA's specialized AI agents}
to synthesize and explore thin-film libraries with lateral-gradient laser spike annealing (lg-LSA)~\cite{bell_lateral_2016}: 
an internal (highest frequency) autonomous loop iteratively proposes optimized property measurements of a given lg-LSA stripe using a hierarchy of optical characterization techniques, while an external autonomous loop proposes and executes the next lg-LSA synthesis via a model that aggregates knowledge obtained by inner-loop iterations.

SARA's nested synthesis, microscopy imaging, and reflectance spectroscopy loops driven by the specialized AIs with AL reflect the hierarchical nature of scientific discovery. 
A primary goal of studying PCSP relationships is the enumeration of all possible syntheses that yield unique materials, a knowledge base that must be built from synthesis phase diagrams over a broad range of synthesis techniques, multiple parameter spaces defined within each technique, and many experimental campaigns to map synthesis phase diagrams in those spaces.  Coordination among the levels of hierarchy is critical for maximizing high-level knowledge generation from low-level experiments, which guides our development of nested AL algorithms that seamlessly incorporate task coordination and uncertainty propagation. This framework is extensible with respect to incorporation of additional levels of hierarchy and/or expansion of techniques, such as additional property measurements and on-the-fly quantum mechanical calculations~\cite{cerqueira_materials_2015,kirklin_open_2015}, that enrich the knowledge within a given level of hierarchy. Networking of capabilities and knowledge sources elevates the use of AI and AL from process optimization to accelerated scientific discovery, a grand vision of AI-assisted science.

\section{Results}

Our goal is to explore synthesis phase diagrams, especially the relatively unexplored ultrafast-annealing region where metastable polymorphs of metal oxides are more likely to form.
Such metastable oxide materials often exhibit improved properties over thermodynamic ambient ground states, and are relevant for countless industrial applications.
The cubic high-temperature polymorph of \ce{ZrO2}, for example, is frequently used as a thermal coating material~\cite{clarke_thermal_2005,hannink_transformation_2000,clarke_materials_2003} due to its low thermal conductivity, while the anatase phase of \ce{TiO2} has attracted interest as a photocatalytic material~\cite{satoh_metastability_2013,cui_first-principles_2016,vu_anataserutile_2012}. These are only two of the most prominent examples of materials systems where metastable phases outperform their respective equilibrium counterparts.

Here we study the Bi--O system, which exhibits a rich phase diagram with dozens of experimentally observed polymorphs. In particular, we focus on the \ce{Bi2O3} composition, for which five distinct crystalline phases are known~\cite{harwig_structure_1978,cornei_new_2006}. 
The monoclinic $\alpha$-\ce{Bi2O3} is the thermodynamic ground state at room temperature, while four high-temperature phases have been reported: tetragonal $\beta$-\ce{Bi2O3}, bcc $\gamma$-\ce{Bi2O3}, cubic $\delta$-\ce{Bi2O3}, and orthorhombic  $\epsilon$-\ce{Bi2O3}. 
The metastable $\delta$-phase has attracted interest as a solid oxide electrolyte in fuel cells~\cite{shuk_oxide_1996}: due to its defective fluorite-type crystal structure with a high concentration of oxygen-vacancies, $\delta$-\ce{Bi2O3} has the highest oxygen ion conductivity of any solid oxide known to date. 
Unfortunately, it exhibits only a narrow thermodynamic stability window between $727 - 824^\circ$C, which has so far precluded its use on an industrial scale. 
Substitution of yttrium or rare earth oxides can stabilize $\delta$-\ce{Bi2O3} to room temperature, but leads to a degraded ion conductivity. 
Hence, efforts have been aimed at finding routes to retain phase-pure $\delta$-\ce{Bi2O3} to ambient conditions~\cite{bell_room_2019}.

\begin{figure*}
    \centering
    \includegraphics[width=0.9\textwidth]{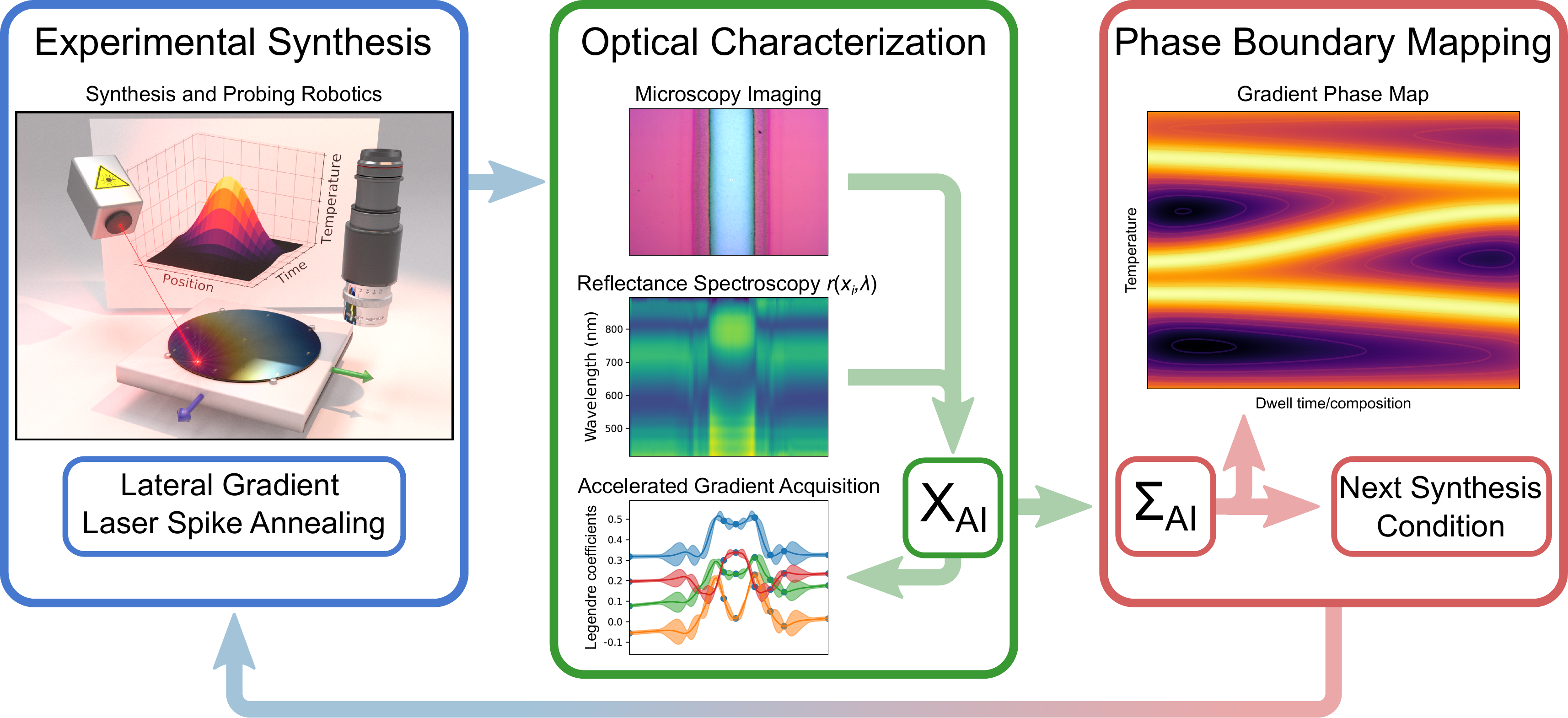}
    \caption{SARA's closed-loop autonomous materials synthesis and discovery cycle. 
    Starting from a set of initially selected processing conditions,
    SARA synthesizes an lg-LSA stripe on a thin film library, and subsequently sends it to its characterization AI agent, \chiai. A schematic illustration of the lg-LSA/camera setup is shown in the left panel with the bi-Gaussian power profile in the backdrop, the laser to the left, a camera to the right, and the thin-film sample mounted on a stage. Using a hierarchy of characterization techniques, 
    \chiai \ analyzes the stripe to determine intricate changes in its optical properties. In particular, \chiai \ first acquires a microscopy image to determine the positions of likely phase boundaries, which informs the reflectance spectroscopy measurements. \chiai's physics-informed AL model accelerates the spectroscopy acquisition, resulting in an accurate gradient model of the lg-LSA stripe. A microscopy image, the reflectance spectroscopy heat map, and the first four Legendre coefficients from the \chiai \ representation are shown in the center panel for a representative lg-LSA stripe. Finally, the gradients are fed into SARA's synthesis AI agent, \sai, which generates a gradient phase boundary map, and also proposes the next experimental processing conditions to improve the phase boundary with as few experiments as possible. A model gradient phase map showing high-gradient regions in yellow is presented in the right panel.}
    \label{fig:overview}
\end{figure*}

Our samples are deposited as amorphous thin films by reactive sputtering on a silicon substrate. For other materials systems, composition spreads can be similarly deposited, allowing the mapping of a composition gradient $c(\mathbf{ x})$ to the location $\mathbf{ x}$ on the substrate.
We process the thin film libraries using lg-LSA to form and kinetically trap metastable phases during the quench to ambient conditions. In contrast to conventional methods for annealing thin film samples, such as hot plate, furnace, and rapid thermal annealing~\cite{Borisenko1997}, lg-LSA allows a controlled and rapid thermal processing over a wide range of conditions in a spatially confined region of less than 1~mm, with quench rates of $10^4-10^7$~K/s
and peak temperatures $T_p$ up to 1400~$^\circ$C
(limited by melt of the silicon substrates).
Scanning a laser beam with a bi-Gaussian-like power profile (see the backdrop in the left panel of Fig.~\ref{fig:overview}) over the film allows a single lg-LSA stripe to produce a spatially inhomogeneous thermal profile $T_{T_p, \tau}(x)$ (where $x$ runs across the stripe).
The duration of heating is characterized by a dwell time $\tau$ defined by the ratio of the laser full-width-half-maximum divided by the scan velocity of the laser (typical dwells range from 10-10,000 $\mu$s).
Hence, at a given dwell time $\tau$, a single lg-LSA experiment produces a  continuous range of temperature conditions wherein phase transitions, including formation of the sought metastable phases, need to be detected with a speed and level of automation commensurate with this robotic synthesis procedure 
in order to fully capitalize and
elevate high throughput synthesis to high throughput discovery of phase boundaries.

In order to reduce both computational and experimental cost,
we need to autonomously map out the processing phase space \{$\mathbf{ x}, \tau, T_p$\} with as few synthesis experiments and property measurements as possible. 
Since the lg-LSA is an irreversible method, a specific position $\mathbf{ x}$ (and potentially its associated composition $c(\mathbf{ x})$ in the presence of a composition gradient) can only be annealed once, further emphasizing the need for optimizing the selection of the processing conditions. 
Once an lg-LSA stripe is processed, a conclusive structural characterization across the thermal gradient is possible with grazing-incidence high-intensity X-ray diffraction (XRD) to resolve the crystal structure.\cite{sutherland_optical_2020} However, access to synchrotron facilities capable of producing X-rays with appropriate wavelength, intensity, and $\mu$m-scale spatial resolution comprise an inherently limited resource that motivates development of alternative phase-boundary-detection methods. 
To address this issue, we developed a complementary technique based on microscopy imaging and optical spectroscopy to rapidly assess phase boundaries.
We recently demonstrated that structural phase changes are directly associated with changes in the optical thin film properties of transparent films, in particular the optical thickness $nd$~\cite{sutherland_optical_2020}, where $n$ is the refractive index and $d$ is the film thickness. 
Essentially, the \textit{gradients} of the optical measurements across an lg-LSA stripe provide a means to map out phase boundaries without explicit crystallographic phase identification, thereby producing an unlabeled processing phase diagram
without expensive XRD experiments.

Herein, we put forth how SARA integrates lg-LSA synthesis and optical phase boundary detection in a hierarchical autonomous workflow 
by employing characterization and synthesis agents, \chiai \ (pronounced Chi AI) and \sai \ (pronounced Sigma AI), respectively, as illustrated in Fig.~\ref{fig:overview}.
Starting with an initial processing condition, SARA synthesizes an lg-LSA stripe on a thin film library.
Then, SARA employs its internal characterization agent \chiai \ to probe the stripe using a set of optical techniques: (a) microscopy imaging to
rapidly inspect 
the anneal stripe (see top panel in ``Optical Characterization'' in Fig.~\ref{fig:overview}),
and (b) more elaborate, but costly, reflectance measurements (see ``Reflectance Spectroscopy'' in Fig.~\ref{fig:overview}).
In particular, \chiai \ uses the observed features from the micrograph as prior knowledge to guide and acquire an accurate reflectance map with as few measurements as possible.
The \textit{gradients} of the reflectance map are then fed into SARA's synthesis AI agent \sai, which incorporates the reflectance gradient information of each lg-LSA stripe into a phase boundary map as a function of the parameters \{$\mathbf{ x}, \tau, T$\}. 
The high-gradient regions of this map determine the boundaries between phase fields, and produce an unlabeled processing phase diagram (see ``Phase Boundary Mapping'' in Fig.~\ref{fig:overview}). 
\sai \ is also responsible for proposing the next, most promising 
synthesis
conditions in order to effectively explore the search space.
We discuss \chiai \ and \sai \ in detail below.

\subsection{\chiai: Accelerating data acquisition and characterization}

\begin{figure*}[th!]
    \centering
    \includegraphics[width=0.9\textwidth]{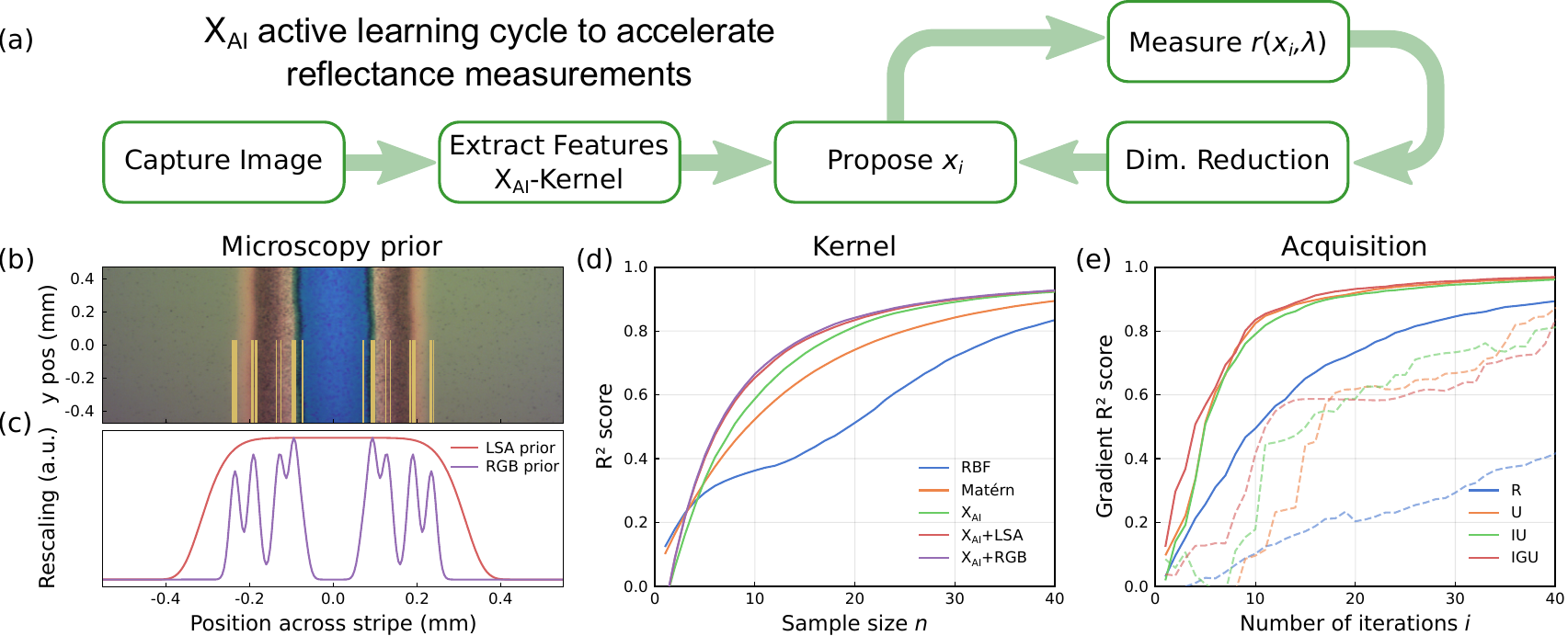}
    \caption{The characterization AL loop to accelerate acquisition of the reflectance spectroscopy necessary for phase boundary detection. Panel (a) illustrates the overall workflow. A microscopy image (b) is captured to extract the stripe features, which are fed as a scaling function to the \chiai \ kernel. The core features are the LSA prior and the RGB transition prior which are sums of generalized Gaussian functions, as shown in panel (c). 
    The corresponding gradient peak positions are denoted as yellow vertical lines in (b). 
     \chiai \ takes these functions as prior knowledge to set up a stripe-specific kernel that facilitates rapid model convergence. 
    The AL loop is performed iteratively on reflectance measurements $r(x_i,\lambda)$ over positions $x_i$, which are expanded into Legendre polynomials to reduce the dimensionality (see center panel in Fig.~\ref{fig:overview}). Panel (d) shows the performance of different kernel designs, illustrating that our \chiai \ kernel with both LSA (\chiai+LSA) and LSA+RGB (\chiai+RGB) priors outperform other, conventional kernels. The performances of the different acquisition function are shown in panel (e) (R: random, U: uncertainty, IU: integratd uncertainty, IGU: integrated gradient uncertainty sampling). The solid lines represent the \chiai+RGB kernel, while the dashed lines correspond to the RBF kernel.}
    \label{fig:iAI}
\end{figure*}

\chiai's primary task is to construct an accurate reflectance spectroscopy map $r(x,\lambda)$ of an lg-LSA sample annealed at $T_p$ and $\tau$ while measuring it at as few positions $x_i$ across the stripe as possible. Since the acquisition time for a single such measurement $r(x_i, \cdot)$ is around 4.5~s, an exhaustive scan across a stripe of 1.5~mm in $10~\mu\text{m}$ intervals requires more than 11~min,
forming one of the main bottlenecks of our HT experimental setup.
To accelerate the reflectance data acquisition,
we propose an AL scheme 
that takes advantage of multimodal measurements
and incorporates physical structure 
into a Gaussian process (GP) regression model to yield highly optimized data acquisition and analysis.

The overall workflow of the \chiai \ cycle is outlined in Fig.~\ref{fig:iAI}~(a). 
In a first step, SARA captures a microscopy image of an lg-LSA stripe to analyze the overall condition of the anneal and to extract key features.
This single RGB image of a stripe is inherently throughput-matched to the lg-LSA synthesis, producing prior knowledge for the \chiai's    
AL cycle to accelerate reflectance measurements.
A representative microscopy image is shown in Fig.~\ref{fig:iAI}~(b). 
Such micrographs can be used to rapidly assess the conditions and the integrity of the anneal. 
Obvious damage of the thin film such as delamination, scratches, or contamination such as dust particles, residual lithography artifacts, and dirt can be easily detected, which invalidates the lg-LSA stripe and can trigger re-synthesis. 
The incorporation of such automated quality control in the autonomous loop alleviates responsibility for the \chiai \ loop to effectively respond to invalid data, a critical aspect of autonomous workflows for robust operation.~\cite{sstein_progress_2019}

SARA proceeds by constructing a stripe-specific GP kernel that incorporates the underlying physics of both the lg-LSA and optical spectroscopy processes. Notably, the bi-Gaussian power profile produces stripes of nearly perfect lateral symmetry at steady state, with their centers reaching the corresponding peak temperatures $T_p$ and the continuous variation in lateral thermal gradient mirrored on each side of the stripe.
We incorporate this structure into the kernel of \chiai \ by forcing its main component to be symmetric around the center of a stripe (see Sec. \ref{sec:iAImethods}).
Additionally, SARA extracts key features of the stripe texture from the micrograph to further improve the kernel design, i.e., by identifying the stripe center, and by detecting systematic optical changes across the stripe that we associate with structural transitions~\cite{sutherland_optical_2020}.
These optical transitions are identified by peaks in the gradient signal across a stripe, the locations of which are shown as vertical yellow lines in Fig.~\ref{fig:iAI}~(b).
Furthermore, the two outermost detected peaks in the gradient signal give an estimate of how wide the lg-LSA stripe is, i.e., where the unannealed, amorphous film ends and the crystallization begins. 
We use slightly broadened peaks in the RGB gradient signal (purple line in Fig.~\ref{fig:iAI}~(c)) and the overall width of the lg-LSA stripe  (red line in Fig.~\ref{fig:iAI}~(c)) as the 
RGB and LSA prior, respectively. 
These two functions 
are then used to rescale the kernel of the GP
in the \chiai \ cycle. Finally, we account for small thickness variations of the film across the stripe by adding a linear component to the kernel.

To improve the efficiency of \chiai, the reflectance $r(x,\lambda)$ at any position $x$ is expanded in Legendre polynomials as a function of wavelength $\lambda$ before it is fed into the GP. 
Since the reflectance varies smoothly with $\lambda$, the Legendre expansion can be truncated between the \nth{10} and \nth{20} order at essentially no loss in accuracy \cite{wang2012legendre} (see Fig.~S1 in the Supplemental Materials~\cite{SupplementalMaterials}), which reduces the dimensionality from 
the 2046 measured photon wavelength
to a compact space of 10--20 Legendre coefficients. 
For our system, we use 16 coefficients throughout.
Figure \ref{fig:overview} (bottom middle) shows the first four Legendre coefficients of the reflectance data and our GP model's posterior predictive mean and uncertainty for those coefficients.

To demonstrate the advantage of our specialized \chiai \ kernel with respect to a set of conventional kernels, we perform statistical benchmarks on 617 lg-LSA experiments at distinct conditions, $(T_p, \tau)_j$.
For each of the stripes, we measure the reflectance at $n$ randomly selected positions $\{x_i\}_j^n$ on a grid spaced $10~\mu\text{m}$ apart, and use these measurements 
as inputs to a GP model with different kernels.
The ground truth is exhaustively measured across the whole stripe ranging over 1.5~mm, corresponding to a total of 151 measurements. 
For a range of $n$,
we repeat this test 32 times for every stripe
with independent random locations 
and average the coefficient of determination $R^2$
for each kernel on the exhaustive data.
This reduces the statistical noise in the results to a negligible value.
Further, we benchmark every kernel with a range of length scales and select the best in terms of $R^2$ score (see Sec.~\ref{sec:metrics}).
By construction, our benchmark disentangles the effects of AL and kernel design, and the kernel with the right inductive bias will express the data best, 
even if all measurements are random.
The results of this benchmark are illustrated in Fig.~\ref{fig:iAI}~(d), showing the performance of the various kernels with respect to the number of random measurements $n$.
The radial basis function (RBF) kernel performs poorly, barely reaching an $R^2$ score of 0.8 within 37 measurements. The Mat\'{e}rn kernel performs better, requiring $n=25$ to reach the same score. 
The \chiai \ kernels perform best: 
depending on whether or not prior knowledge from the microscopy image is included (``\chiai+LSA'' with LSA prior only, and ``\chiai+RGB'' with LSA and gradient peak prior), we obtain an $R^2$ value of 0.8 with as few as 16 sampling points. 
The precise modeling of the optical measurements, its incorporation into the AL model, 
and the model's initialization with the RGB-image prior knowledge all contribute to the fast learning rate at the onset of autonomous experimentation, as required for efficient AL.

Having designed the kernel for the \chiai \ cycle,
we turn our attention to the acquisition function,
that is, the function that chooses the next measurement based on the available information.
An important component of many performant acquisition functions is the reduction of uncertainty in a target variable.
Here, we benchmark three different acquisition functions, two of which are non-standard.
In particular, we study (U) uncertainty sampling, which chooses the next measurement at the point of maximum uncertainty in the Legendre coefficients; 
(IU) integrated uncertainty sampling, which selects the point that minimizes the integrated uncertainty over the whole sampling domain; 
and (IGU) integrated gradient uncertainty sampling, which is similar to IU, but reduces the overall uncertainty in the gradients of the model.
Importantly, 
the last strategy targets our quantity of interest, 
since the reflectance gradients are indicative of the phase boundaries in the processing phase diagram.
For this reason, we quantify the error of the model in the gradients, rather than the error to the observed data. 
Since we cannot directly observe the gradients,
we generate ground truth data by training our GP model on the exhaustive measurements and taking the derivative of the fitted model. 
We then record the $R^2$ score of the derivatives of the model for each of the acquisition functions at every iteration.

In Fig.~\ref{fig:iAI}~(e) we show the performance of the various sampling strategies as a function of AL iteration $i$, using either the \chiai+RBG kernel (solid lines) or the RBF kernel (dashed lines).
The best performance is achieved with the stripe-specific, highly optimized \chiai+RBG kernel in conjunction with IGU sampling, reaching an $R^2$ score of 0.8 and 0.9 within 9 and 15 iterations, respectively.
Note that random sampling with the best kernel design still outperforms the best sampling strategies with the worst kernel.
Further, the acquisition functions do not differ markedly with the \chiai+RGB kernel, 
highlighting the importance of incorporating the problem structure into our AI model and AL cycle.
Compared to random sampling with an RBF kernel,
the best strategy accelerates the acquisition and characterization by a factor of 
9.7 
for an $R^2$ of 0.8, 
approximately one order of magnitude.

\subsection{\sai: Accelerating phase exploration and processing conditions}
\begin{figure*}[th!]
    \centering
    \includegraphics[width=0.9\textwidth]{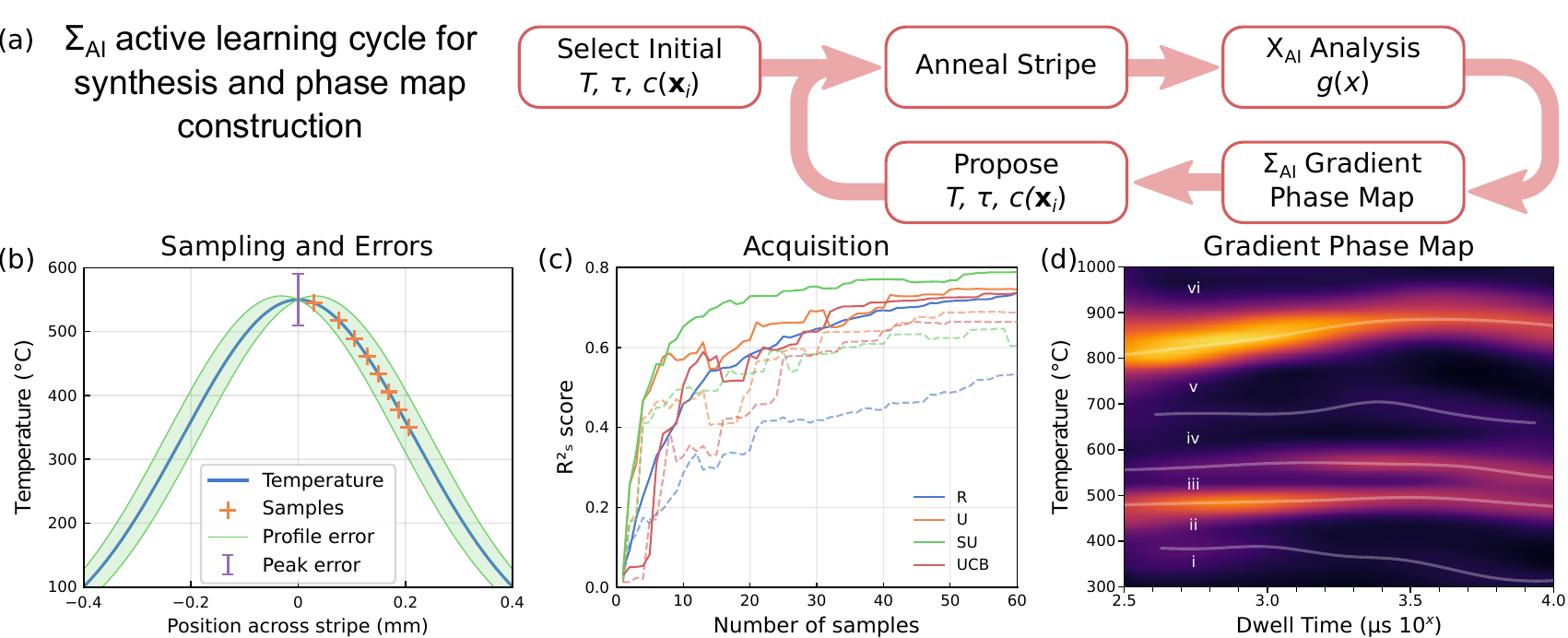}
    \caption{The synthesis AL loop to accelerate materials exploration. Panel (a) illustrates the overall external workflow. Starting from an initial set of conditions, an lg-LSA stripe is annealed and processed by \chiai. The gradients are then fed into the \sai \ agent, which constructs a (preliminary) gradient phase map and proposes the next experimental conditions. The transformation of the \chiai \ reflectance gradients requires a rigorous error assessment and propagation, as shown in panel (b). Due to the symmetric \chiai \ kernel, only one side from the stripe center is sampled on a uniform temperature mesh. The errors propagated to the \sai \ stem from the variation in the peak temperature (peak error) and the gradient of the temperature profile (profile error). Panel (c) shows the performance of different \sai \ acquisition strategies: random (R), uncertainty (U), stripe uncertainty (SU), and upper confidence bound (UCB) sampling. The solid and dashed lines correspond to GP regression with and without input uncertainty, respectively. Panel (d) shows the gradient phase map of \ce{Bi2O3}, where the peak ridges are highlighted with light lines. The phase regions are labeled \textit{a posteriori} with selected XRD measurements, from low to high temperatures: amorphous as-deposited ($i$), rearranged, densified amorphous ($ii$), $\delta$-\ce{Bi2O3} ($iii$), mixed-phase region of $\delta$ and $\beta$-\ce{Bi2O3} ($iv$), pure $\beta$-\ce{Bi2O3} ($v$), and, finally, melt-quenched amorphous ($vi$).}
    \label{fig:eAI}
\end{figure*}

Once an lg-LSA stripe has been processed by \chiai, its output reflectance gradient information is fed into the external synthesis AI agent, \sai.
Its main task is threefold: (i) assemble the incoming data, (ii) propagate uncertainty from every lg-LSA experiment to predict the gradient signal and its uncertainty throughout the search space, and (iii) ultimately propose new conditions for the synthesis experiments.
The overall workflow of this process, which integrates the techniques described below, is shown in Fig.~\ref{fig:eAI}~(a).

The optical data of an lg-LSA anneal is processed through the nested \chiai \ loop, the output of which is the gradients of the reflectance spectroscopy across a stripe, 
$g(x)= \| \partial_x r(x, \cdot) \|_2$.
This spatial gradient information is then transformed onto a temperature scale 
based on the Gaussian-type temperature profile 
$T_{T_p, \tau}(x)$
shown in Fig.~\ref{fig:eAI}~(b) (blue line).
Since the \chiai \ kernel is symmetric up to the linear term, the gradient information is symmetric about the peak temperature $T_p$,
so that we only need to sample $g(x)$ along one side from the stripe center (orange crosses in Fig.~\ref{fig:eAI}~(b)).

In principle, one single lg-LSA stripe would produce the complete temperature conditions between room temperature $T_r$ and $T_p$ at a given dwelltime $\tau$. 
Hence, the set of metastable materials and their transition conditions would be available from a single stripe if one selected a high $T_p$ (e.g., $1400^\circ$C) and $T_\text{min}=T_r$. 
In practice, the concomitant increase in temperature gradient with $T_p$ would require progressively higher spatial resolution to characterize the full range of transitions and results in undesirably high uncertainty in the modelled temperature.   
With our experimental characterization technique, the spatial resolution is limited to approx.~$10~\mu$m, and thus the design of lg-LSA synthesis conditions must be done under consideration of the position-dependent temperature variation within a single spectroscopy measurement, which makes the selection of $T_p$ at a given $\tau$ a non-trivial decision based on the aggregate information that can be gained from the entire lg-LSA stripe.

Properly propagating the multiple sources of uncertainty from synthesis and characterization through the model of the phase boundary map is extremely important: 
in standard GP regression, the inputs are assumed to be free of noise, but accounting for such errors is crucial when dealing with experimental measurements. 
Here, we include and propagate the uncertainties of the inputs due to two sources.
First, the peak temperature reached in an lg-LSA anneal can vary within an error range of up to $\sigma_{T_p} = 25^{\circ} \text{C}$ at 1400~$^\circ$C due to fluctuation in the laser power, even after reaching steady state (peak error in Fig.~\ref{fig:eAI}~(b)).
Second, the temperature profile itself gives rise to an error proportional to the spatial rate of change $\sigma_T(x) \propto |\partial_x T_{T_p, \tau}(x)|$,
as shown by the green area in Fig.~\ref{fig:eAI}~(b). 
Note that the error bars in Fig.~\ref{fig:eAI}~(b) are not to scale and intended solely for a schematic illustration.

As opposed to the \chiai \ model of optical spectroscopy, the gradient map in synthesis space has no analogous physics-based model, in part because too few synthesis phase diagrams are known, and their underlying features remain an open question. The large dynamic range of dwell time motivates its logarithmic sampling, and the distinct influence of temperature and dwell time on synthesis motivate independent parameterization of these two dimensions. 
While we aim to learn more structured representations of synthesis phase diagrams in future refinements of the SARA framework, 
for the purposes of the present work,
we find a Mat\'{e}rn kernel,
with separate length scales for the temperature and dwell time dimensions, 
to enable rapid model convergence in the \sai \ loop while remaining flexible with respect to the gradient map in synthesis phase space.

In contrast to the \chiai \ cycle, 
there is more opportunity to incorporate structure based on prior knowledge into the acquisition function, rather than the kernel, of \sai.
As shown in Fig.~\ref{fig:eAI}~(c), random sampling performs only slightly worse than more sophisticated acquisition methods like uncertainty sampling or upper confidence bound (UCB) sampling
in terms of $R^2_s$, a generalization of $R^2$ that takes into account the heteroscedasticity of the data due to the propagation of uncertainty (see Sec. \ref{sec:metrics}).
This behavior can be understood by considering the following: every experiment at $\{T_p, \tau\}$ produces a range of temperatures $T<T_p$ at which new information is obtained, thereby reducing the uncertainty not only at $T_p$, but in a wide range of temperatures below it. 
Hence, uncertainty sampling at $T_p$ and $\tau$ alone is a poor strategy. 
To address this issue, we introduce stripe uncertainty (SU) sampling, which takes into account the uncertainty in the \textit{whole temperature range} between $T_\text{min}$ and $T_p$. 
{This strategy greatly improves performance, reaching  $R^2_s >0.7$   within 15 iterations.}

\begin{figure*}
    \centering
    \includegraphics[width=0.9\textwidth]{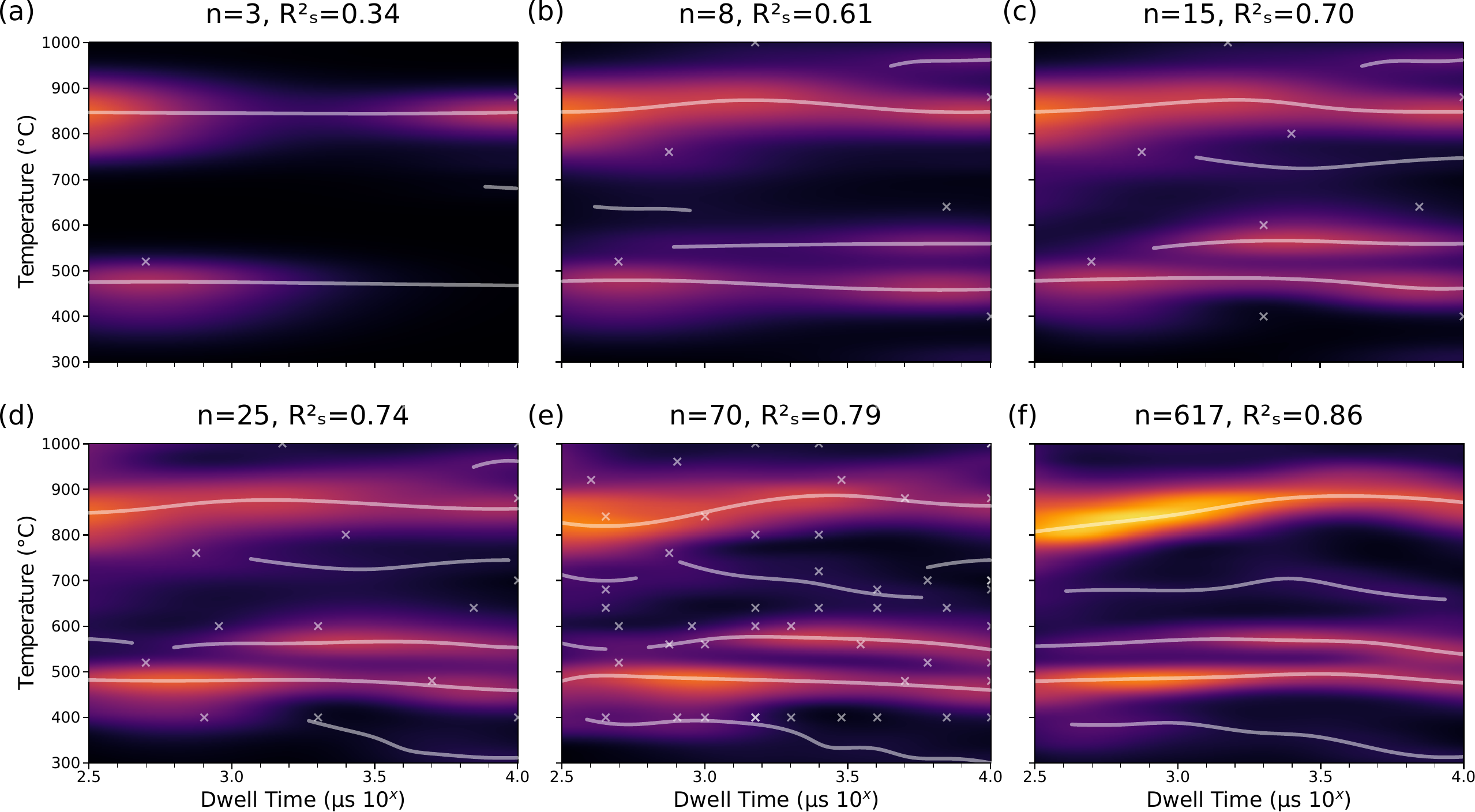}
    \caption{The evolution of the actively learned gradient phase map of the \ce{Bi2O3} system at selected number of iterations $n$. We use the stripe uncertainty (SU) acquisition strategy, starting from a randomly selected condition $(T_p,\tau)_1$. The gradient ridges are shown as white lines, and the conditions  $(T_p,\tau)_i$ at which the experiments have been performed are shown as white crosses (note that not all crosses are shown, since the plots have been cropped to a smaller range than the range of experimental conditions). For each panel, the number of sampled conditions $n$ is indicated at the top, together with the corresponding $R^2_s$ score. The final, exhaustively sampled phase map is shown in panel (f).}
    \label{fig:evolution}
\end{figure*}

The plot in Fig.~\ref{fig:eAI}~(d) shows the gradient heat map of \ce{Bi2O3} from an exhaustive sampling of all 617 lg-LSA stripes, with gradient peaks at every value of $\tau$ highlighted in white. 
Note that these peaks are connected and form ridges, that can be well interpreted as phase field boundaries. 
To label these phase fields, we selectively collect and analyze XRD data of lg-LSA stripes annealed at conditions close to the field centers (see Supplemental Materials for details~\cite{SupplementalMaterials}).
Notably, only few XRD measurements suffice to label the phase map, once the phase boundaries have been determined via the reflectance data.
The phase field $i$ below approx.~$350^\circ$~C corresponds to the as-deposited amorphous film, while a slight gradient ridge separates it from $ii$, a densified, amorphous regime.
At approx.~$500^\circ$~C there is a ridge that extends across the complete dwell range, corresponding to the crystallization onset of the  $\delta$-phase in domain $iii$. 
The boundary separating $iii$ from $iv$ at 
approx.~$550^\circ$~C corresponds to the onset of a two-phase region, where both the $\delta$ and $\beta$-phase of \ce{Bi2O3} coexist, and above approx.~$650^\circ$~C we observe phase-pure $\beta$-\ce{Bi2O3} in $v$. 
The phase field above approx.~$810^\circ$~C corresponds to amorphous \ce{Bi2O3} that reforms after quenching from melt (the bulk melting temperature of \ce{Bi2O3} is $817^\circ$C~\cite{madelung_bismuth_1998}) and stretches out across all values of $\tau$.

A representative evolution of the actively learned gradient phase map is shown in Fig.~\ref{fig:evolution}, with six snapshots in the panels from (a) to (f). Panel (a) shows the preliminary gradient map at iteration $n=3$: the two gradient ridges spanning all dwells qualitatively correspond to the crystallization  boundaries either from melt (($v$--$vi$) in Fig.~\ref{fig:eAI}~(d)) and the deposited, densified thin film ($ii$--$iii$). At $n=8$ in panel (b), we detect the onset of the two-phase region ($iii$--$iv$), and at $n=15$ (panel (c)) the phase-pure $\beta$-\ce{Bi2O3} boundary ($iv$--$v$). With only $n=25$ iterations we identify the last boundary, namely the amorphous densification onset ($i$--$ii$, see panel (d)). At this point, the overall features of the gradient phase map are already qualitatively captured completely, and subsequent iterations merely refine the boundary locations (panel (e)), getting closer to the exhaustive phase map in panel (f).

Two factors are crucial for \sai \ to achieve a factor of approx. 14 acceleration to reach $R^2_s = 0.7$ compared to random sampling without propagation of input uncertainties.
First, incorporating materials synthesis into our SARA discovery framework allows to check for convergence of the phase diagram on the fly. 
Even with random sampling, the possibility of quantifying the progress and monitoring convergence in the gradient mapping
informs us how well the phase space has been sampled, thereby significantly reducing the resource cost. 
Second, the comprehensive uncertainty propagation in conjunction with the stripe uncertainty acquisition function realizes the full potential of AI and AL and decreases the required samples to a fraction of the exhaustive measurements.
Importantly, SARA's overall AL acceleration is the {\it product} of the acceleration factors of
 \chiai \ and \sai, due to the cycles' nested design.

\section{Discussion}
In conclusion, we have developed  SARA, an AI-driven autonomous closed-loop materials discovery framework that integrates robotic materials synthesis with automated microscopy imaging and reflectance spectroscopy characterization.
 SARA incorporates a set of nested AL loops based on specialized physics-inspired  Gaussian process regression models to synthesize, characterize, and iteratively explore non-equilibrium synthesis phase maps using high-throughput lg-LSA thin-film processing.
In particular, {
SARA tightly integrates the physics of the experiments and quantifies} experimental uncertainties in both the inputs and the outputs of the model.
We highlight { 
SARA's capabilities  on the \ce{Bi2O3} system 
by showing that SARA reduces the time to map the system’s phase boundaries by more than two orders of magnitude, in contrast to random or exhaustive searches.
In particular, SARA identifies the synthesis conditions that trap metastable $\delta$-\ce{Bi2O3} at room temperature, a promising solid oxide electrolyte for fuel cell applications.}

This speedup in synthesis and data-acquisition is a fundamental prerequisite for paving the path towards exploratory HT efforts with additional chemical degrees of freedom, extended processing parameters, and when targeting property optimization. The gradient phase map construction can be extended to additional degrees of freedom, e.g., on composition spreads over a continuous range of chemistries. {
SARA's nested AI architecture also allows the incorporation of additional agents for multi-objective optimization efforts by including robotic measurements of target properties.}
In addition to phase boundary mapping, research objectives for which SARA would enable new modalities of materials design include: discovery of a synthesis condition for a not-yet-synthesized phase, extension of the optical spectroscopy to characterize visible absorption to identify syntheses of materials for solar energy applications, and incorporation of new performance characterization such as electrical conductivity measurements. These latter examples involve mapping of synthesis phase diagrams in the context of performance metrics for a target application, the central goal of studying PCSP relationships. SARA's autonomous execution of such studies constitutes a grand vision of AI-assisted materials science.

\section{Materials and Methods}
\subsection{Experiments and measurements}
\subsubsection{Thin film deposition}
We used thermally oxidized (200~nm oxide), highly doped (\emph{p}-type, 0.01--0.02~$\Omega$ cm), Si wafers with lithographically patterned gold alignment marks as substrates for our thin film deposition. RF reactive sputtering from a \ce{Bi} target in an atmosphere of 8~mTorr Ar and 2~mTorr \ce{O2} was employed to deposit the \ce{Bi2O3} sample in a custom built sputter system. The substrate was rotated while operating the target at an RF power of 20~W to create a 170~nm thick film with $<10\%$ thickness variation.

\subsubsection{Lateral gradient laser spike annealing}
The lg-LSA anneals were conducted using a CW \ce{CO2} laser operating at $\lambda = 10.6\ \mu$m and maximum power of 125~W which was configured to produce a power profile with a bi-Gaussian shape ($320~\mu$m-wide lateral FWHM and $80~\mu$m-long longitudinal FWHM). To reach steady-state, each anneal was conducted on a 5~mm long stripe, with peak temperatures ranging from 400$^{\circ}$C to 1300$^{\circ}$C and processing dwell times between 250~$\mu$s and 10~ms. 
The stripes were located 2~mm apart to avoid thermal overlap between anneals. 
With this configuration, a 100~mm diameter wafer offers space for a total of up to 625 stripes with distinct anneal conditions. 
Note that the dwell $\tau$ is related to the scan velocity $v$ via the FWHM of the laser in the scan direction (longitudinal) through $\tau=\frac{\text{FWHM}}{v}$. $\tau$ is approximately the time scale during which the temperature is within 5~\% of the peak temperature~\cite{bell_lateral_2016}. To avoid potential location bias on the wafer arising from variations in film thickness, the anneal locations were randomized across the thin film with respect to the $T_p$ and $\tau$. In total, we annealed 617 lg-LSA stripes on our \ce{Bi2O3} sample with $400 \leq T_p \leq 1300^\circ$C, and $250 \leq T_p \leq 10,000\mu$s.

\subsubsection{Microscopy imaging}
We used a Thorlabs CMOS camera (RGB channels with $1024 \times 1280$ pixels), which was aligned normal to the sample, together with a coaxial illumination using white light over a spot size of approximately 1~mm in diameter. The camera magnification was set to produce a field of view (FOV) of approximately 1~mm horizontally, resulting in a spacing of 0.92~$\mu$m between pixels.

\subsubsection{Reflectance spectroscopy}
A white light source ($400 <\lambda < 900$~nm) was focused down to a single 10~$\mu$m diameter spot using optical fibers to locally illuminate the sample, allowing spatially resolved reflectance measurements. We used a Flame Spectrometer from Ocean Optics to collect the reflectance spectroscopy with an optimized integration time of $\approx 4500$~ms. The reflected light was measured from $\lambda=340$ to $\lambda < 1026$~nm at 2046 discrete values. The reflectance data was calibrated and normalized with respect to a dark reference spectrum, and a spectrum from an Ag-coated mirror. For the exhaustive reflectance measurements, the optical fiber was scanned across an lg-LSA stripe over a range of 1.5~mm in 10~$\mu$m increments, leading to 151 samples per stripe.

\subsubsection{X-ray diffraction}
The XRD data was collected using the ID3B beamline at the Cornell High Energy Synchrotron Source (CHESS) with a 9.7~keV beam, which was focused to a spot size on the sample of $20\ \mu\text{m}\times40\ \mu\text{m}$ at a 2$^\circ$ angle of incidence. A Pilatus 300K detector was used to capture the diffracted signal. The XRD data were collected every 10~$\mu$m across a stripe with a 50~ms integration time for each frame. The 2D detector data were integrated along the $\chi$ direction using pyFAI~\cite{Ashiotis2015}.

\subsection{Computational methods}

In the following, bold lower case letters refer to vectors and bold upper case letters refer to matrices.
Given a collection of inputs $\bold X = [\bold x_1, \hdots, \bold x_n]$ of a function $f$,
we let $f(\bold X)$ be the result of the application of $f$ to each column of $\bold X$, $f(\bold X) := [f(\bold x_1), \hdots, f(\bold x_N) ]$.

\subsubsection{Gaussian Processes}

A Gaussian Process (GP) is a distribution over functions
whose finite-dimensional marginal distributions 
are multivariate normal.
That is, for any sample $f$ of a GP, 
and any finite selection of inputs $\bold X$, we have
$
f(\bold X) \sim \N(\boldsymbol \mu_{\bold X}, \boldsymbol \Sigma_{\bold X})$,
for some mean vector $\boldsymbol \mu_{\bold X}$ and covariance matrix $\boldsymbol \Sigma_{\bold X}$.
In fact, analogous to the multivariate case, a GP is completely defined by its first and second moments:
a mean {\it function} $\mu(\cdot)$
and a covariance {\it function} $\kappa(\cdot, \cdot)$,
also known as a kernel.
In particular, if $f \sim \GP(\mu, \kappa)$ 
then for any finite collection of inputs $\bold X$,
\begin{equation}
f(\bold X) \sim \N(\mu(\bold X), \kappa(\bold X, \bold X)),
\end{equation}
where $\kappa(\bold X, \bold X)$ is the matrix whose $(i,j)^\text{nth}$ entry is $\kappa(\bold x_i, \bold x_j)$.
Fortunately, the posterior mean $\mu_p$ and posterior covariance $\kappa_p$ of a GP
conditioned on observations with normally-distributed noise have closed forms
and only require linear algebraic operations:
\begin{equation}
\label{eq:gp_posterior}
    \begin{aligned}
        \mu_p(\bold x_*) &= \mu(\bold x_*) + \kappa(\bold x_*, \bold X) \bold \Sigma_{\bold X}^{-1} (\bold y - \mu(\bold X)),\\
        \kappa_p(\bold x_*, \bold x_*') &= \kappa(\bold x_*, \bold x_*') - \kappa(\bold x_*, \bold X) \bold \Sigma_{\bold X}^{-1} \kappa(\bold X, \bold x_*'),\\
    \end{aligned}
\end{equation}
where for homoscedastic regression,
$\bold \Sigma_{\bold X} = k(\bold X, \bold X) + \sigma_y^2 \bold I$ and $\sigma_y$ is the standard error of the target $\bold y$.
Since a GP's behavior 
is chiefly determined by the kernel, 
its performance can be improved dramatically by incorporating important problem structure into the kernel.
For more background on Gaussian processes, see \cite{rasmussen:2005:gpml}.
For the present work, we developed a Gaussian process framework in Julia \cite{julia2017bezanson}
with which we implemented SARA's active learning technology.

\subsubsection{Active Learning}
The field of active learning considers the problem of selecting data in an optimal way
in order to reduce the total amount of data 
that is required to effectively train a model \cite{cohn1996active, settles_active_2009}.
To this end, the notion of an {\it acquisition function} is important.
An acquisition function $a(\bold X, \bold y)$ dependents on currently available data,
and outputs a suggested observation $\bold x^*$.
For example, if $f | \bold X, \bold y$ denotes the posterior of $f$ after having seen the data,
and var$(f| \bold X, \bold y)$ is the posterior variance (itself a function), then
\begin{equation}
\label{eq:us}
\arg \max_{\bold x^*} \text{var}(f | \bold X, \bold y)(\bold z)
\end{equation}
defines an acquisition strategy known as {\it uncertainty sampling}. 
Other acquisition functions
are based on upper-confidence bounds,
expected improvement, and probability of improvement.
Overall, an important ingredient for active learning is the quantification of uncertainty, which is a strength of Bayesian models. 
In the realm of Bayesian models, 
Gaussian processes are of particular importance 
because of their unique combination of flexibility, closed-form inference formulas, and uncertainty quantification.
For these reasons, we chose to build SARA's computational backbone on Gaussian processes.

\subsubsection{Input Noise}
\label{sec:inputnoise}
Due to the importance of uncertainty quantification for AL,
it is critical to take all sources of uncertainty into account.
In the case of SARA,
it is crucial to not only account for errors in the measurements (i.e., model outputs), 
but also the experimental conditions (i.e., model inputs) due to intrinsic experimental uncertainties in the temperature profile.
However, the general problem of posterior inference with input noise is intractable.
For this reason, 
one needs to employ approximate methods like
variational approximations \cite{titsias2009variational, lazaro2011variational}, 
Markov-chain Monte Carlo \cite{neal1997monte}, 
or methods that transform the problem of homoscedastic regression with input noise to one of heteroscedastic regression \cite{le2005heteroscedastic} without input noise \cite{kersting2007most, dallaire2009learning, snelson2012variable}.
A particularly efficient technique is that of McHutchon and Rasmussen \cite{mchutchon2011input},
which is based on propagating the input uncertainty using a linear approximation of the standard posterior mean.
According to this model,
given the regular posterior mean $\mu_p(x)$,
the input-noise-corrected version can be computed by updating
\begin{equation}
\label{eq:nigp}
   \bold \Sigma_{\bold X} \leftarrow \bold \Sigma_{\bold X} + \text{diag}(\sigma_x(\bold X) \odot \partial_x \mu_p(\bold X))^2 
\end{equation}
in the equations~\eqref{eq:gp_posterior} for the GP posterior.
Notably, we generalize the original work 
in making the input uncertainty $\sigma_x(\bold X)$ dependent on the input.
This is possible because the non-constant uncertainties in SARA's experimental process can be estimated well by physical considerations (see Section \ref{sec:eAImethods} for details).
Lastly, note that Eq.~\eqref{eq:nigp} makes the approximate posterior uncertainty dependent on the values of the data via the posterior mean, not just the locations of the measurements.

\subsubsection{\chiai}
\label{sec:iAImethods}
The goal of the \chiai \ is to infer the reflectance
$r(x, \lambda)$ using the least number of measurement locations $x_i$ as possible.
Each measurement of the inner loop 
acquires the wavelength-dependent spectroscopic reflectance of the underlying thin film,
that is, a vector whose entries correspond to reflectance intensities at a given wavelength. 

To aid the efficiency of our model, we first reduce the dimension of the output 
by projecting it onto the basis of a small number (10 to 20) of Legendre polynomials.
Since the signal is smooth as a function of wavelength, it admits a sparse approximation in this basis, allowing the compression of the signal with virtually no loss of information \cite{wang2012legendre} (see also Supplemental Materials).
The AL cycle then works on the dimensionality-reduced form of the reflectance data.

In the following, we describe the construction of the \chiai \ kernel function, which integrates special structure of the data and is a critical part of the \chiai.
In particular, the kernel incorporates 1) lateral symmetry, 2) variance scaling based on RGB data,
and 3) asymptotically linear behavior.
Starting with a Mat\'ern 5/2 kernel $k$ with a length scale $l$,
we symmetrize it via
$k_{\text{sym}}(x,y) = k(x, y) + k(x-c, y-c)$ around the stripe center $c$,
which we estimate from the RGB images.
We incorporate further information from the RGB images by scaling the kernel with the LSA or RGB prior function $f_{\text{rgb}}$ shown in Figure \ref{fig:iAI} (c).
In particular, 
we use the peaks in the RGB gradient signal and slightly broaden them 
by a Gaussian with $\sigma=20\mu\text{m}$, and sum them to our RBG prior function (purple line in Fig.~\ref{fig:iAI}~(c)). 
Additionally, the overall width of the lg-LSA stripe gives rise to the LSA prior, which is a generalized Gaussian with a wide shape parameter of $\beta = 4$ and a scale parameter $\sigma$ defined by the stripe width (red line in Fig.~\ref{fig:iAI}~(c)). 
$f_{\text{rgb}}$ is then given by a weighted sum of these two prior functions.
This scaling constrains the search space,
since we don't expect a lot of change in the underlying material if the experimental conditions (e.g. temperature) stay similar,
and gives rise to the kernel 
$f_{\text{rgb}}(x) k_{\text{sym}}(x, y) f_{\text{rgb}}(y)$.
Lastly, we incorporate an asymptotically linear behavior, due to thickness variations in the wafer,
with the linear kernel $k_{\text{line}}(x,y) = x \cdot y + b$, where $b$ is a constant that controls the variance of the bias term of the line.
As a result, the \chiai \ kernel for one Legendre coefficient is proportional to
\begin{equation}
k_{\text{\chiai}}(x,y) = f_{\text{rgb}}(x) k_{\text{sym}}(x, y) f_{\text{rgb}}(y) + k_{\text{line}}(x, y).
\end{equation}
For all the Legendre coefficients, 
we then use a GP with the kernel $a_i k_{\text{\chiai}}(x, y)$, where $\{a_i\}$ are scaling coefficients that incorporate the different variances of the Legendre coefficients, to learn the reflectance map.
This can also be interpreted as computing a vector-valued GP with the matrix-valued kernel
$\bold K_{\text{\chiai}}(x,y) = \text{diag}(\bold a) \ k_{\text{\chiai}}(x,y)$,
where $\bold a$ is the vector of scaling coefficients.
For a comprehensive review on matrix-valued kernels, see \cite{alvarez:2012:kvf}.
The length scale $l$ of the Mat\'ern kernel can be optimized via maximization of the marginal likelihood \cite{rasmussen:2005:gpml}.
However, to make the reported results in Fig. \ref{fig:iAI} (d) independent of this non-convex optimization procedure,
we ran the benchmarks using a range of fixed length scales and reported the best performing combination for each kernel.

Regarding the acquisition function,
in addition to uncertainty sampling,
we benchmark the \chiai \ using integrated uncertainty sampling (IU),
a policy that reduces the total variance over a set of potential measurement locations $\bold Z$.
In particular, IU is defined by
\begin{equation}
\label{eq:ius}
    \arg \min_{\bold x^*} \sum_{\bold z \in \bold Z} \text{var}(f | \bold X, \bold y, \bold x^*)(\bold z),
\end{equation}
where $\bold X$ is the set of inputs and $\bold y$ is the set of outputs of the model.
Note that we can calculate the quantity $\text{var}(f | \bold X, \bold y, \bold x^*)$
because the standard posterior GP variance only depends on the measurement location, not the value $y^*$.
Lastly, we note that the derivative of a GP is also a GP \cite{solak2002derivative}.
Plugging the derivative GP into Eq.~\eqref{eq:ius} yields integrated gradient uncertainty sampling (IGU),
which achieves the best performance in the \chiai \ acquisition benchmark (see Fig.~\ref{fig:iAI}).

\subsubsection{\sai}

\label{sec:eAImethods}

The \sai \ works to identify phase regions and their boundaries in the temperature-dwelltime space, and more generally, the processing-composition space.
Importantly, the raw reflectance data cannot be used directly for this task
because of two reasons.
First, the data is measured as a function of position, not temperature.
Therefore, we convert the stripe-specific reflectance function $r_{T_p, \tau}(x, \lambda)$ to the temperature domain using the temperature profile $T_{T_p, \tau}$,
yielding $r_{T_p, \tau}(T, \lambda)$.
Second, the reflectance varies not just with the phase behavior, but also with the film thickness across the wafer.
For this reason, we calculate the $L_2$-norm of the \textit{rate of change} of the spectroscopic reflectance,
which is invariant to linear thickness variations of the film.
In particular, for all $T < T_p$ we want to infer
 \begin{equation}
 \label{eq:d}
 d(T, \tau) := 
 \sqrt{\int \left( \frac{\partial r_{T_p, \tau}(T, \lambda)}{\partial T} \right)^2 \text{d} \lambda}.
 \end{equation}
$d$ quantifies how much the spectroscopic reflectance changes as a function of temperature and dwell time
and is a strong indicator of phase changes~\cite{sutherland_optical_2020}.
Estimating the phase boundaries then reduces to getting an accurate estimate of $d$ over all $(T, \tau)$ (and potentially composition $c$).
This is the goal of the \sai \ loop.
 
Crucially, experimental errors can occur in $x$, and therefore in $T$,
making it imperative to quantify the uncertainty due to these input errors and propagate them to the \sai. 
Indeed, our benchmarks show that ignoring these uncertainties leads to a significant deterioration in active learning performance (see Figure \ref{fig:eAI}(c)).

To this end, we now discuss the intrinsic experimental uncertainties due to the temperature profile $T_{T_p, \tau}(x)$ of the laser. 
In particular, we compute the variance of the true temperature around the value predicted by the temperature profile as a function of position by 
\begin{equation}
\label{eq:temperature_uncertainty}
       \sigma_T^2(x) = \sigma^2_{T_p} \left(
        \frac{T_p}{1400}\right)^2 
        + \sigma_x^2 \ \left(\frac{\partial T_{T_p, \tau}(x)}{\partial x}\right)^2
\end{equation}
where $\sigma_{T_p}$ is the standard error in the peak temperature and
$\sigma_x$ is the standard error in the position.
The first term quantifies the error at the peak temperature
which is largest at high temperatures ($1400^{\circ}$C)
and falls off linearly with $T$.
The second term quantifies uncertainties of the temperature
profile, which are not just due to limited spatial resolution, 
but also encompass random asymmetries in the profile of the laser.
The form of term is derived using standard error propagation techniques \cite{tellinghuisen2001statistical}.
For the results reported herein, $\sigma_{T_p} = 25^{\circ} \text{C}$ and $\sigma_x = 50 \mu\text{m}$.

The expression for the temperature uncertainty in Eq.~\eqref{eq:temperature_uncertainty} is then used 
in conjunction with Eq.~\eqref{eq:nigp} 
to compute a GP that comprehensively quantifies the uncertainties in the Legendre coefficients of the optical reflectance as a function of temperature.
To compute $d$ in Eq.~\eqref{eq:d},
one simply sums the squared derivatives of the GPs of the Legendre coefficients of the reflectance:
\begin{equation}
\label{eq:dfromgp}
    d(T, \tau) = \sqrt{\sum_i (\partial_T \mu^{(i)}_p(T))^2}
\end{equation}
Since we have access to the uncertainties in $\mu_p^{(i)}$ from the GP,
we can use uncertainty propagation techniques on Eq.~\eqref{eq:dfromgp} 
to calculate a first-order uncertainty estimate of $d(T, \tau)$.
For the outer loop, we used a two dimensional Mat\'{e}rn 5/2 kernel with 
different length scales across each dimension.
This allows the GP to learn independent sensitivity parameters
of the experiment for the input dimensions.
Note that the \sai \ benchmarks in Fig. \ref{fig:eAI} (c) were carried out with fixed length scales to disentangle the effects of different acquisition functions and hyper-parameter learning.

For the \sai, we designed an acquisition strategy 
that incorporates the property that a single stripe generates data throughout a range of temperatures.
In particular, given experimental conditions $\bold x_s$ 
that give rise to a stripe ($T_p$, $\tau$, etc.),
we sum the uncertainties of all relevant observations $\bold x_i$
that are in the set Stripe$(\bold x_s)$
of conditions on the stripe $\bold x_s$.
In particular, we propose {\it stripe uncertainty sampling}:
\begin{equation}
    \label{eq:stripe_uncertainty}
    \arg \max_{\bold x_s} \sum_{\bold x_i \in \text{Stripe}(\bold x_s)}
    \text{var}(f | \bold X, \bold y)(\bold x_i).
\end{equation}
Notably, one can use the same principle to generalize 
other acquisition functions.
In fact, we investigated a stripe upper-confidence bound sampling policy.
However, it performed worse or equal to the simpler stripe uncertainty sampling policy above.
The synergy of the comprehensive uncertainty quantification 
and the stripe sampling function yields considerable benefits,
as displayed in Figure \ref{fig:eAI} (c).

\subsubsection{Error Metrics}
\label{sec:metrics}
In our benchmarks of the kernels and acquisition functions for the inner loop, we used the coefficient of determination $R^2$ to measure performance, defined by
\begin{equation}
    R^2 
    = 1 - \frac{\sum_i (f(x_i) - y_i)^2}
    {\sum_i (\mu(\bold y) - y_i)^2},
\end{equation}
where $\mu(\bold y)$ is the mean of the data $\bold y$.
The advantage of using $R^2$ over other canonical measures like the mean-squared error is that it is
dimensionless and easily interpretable as the proportion of the variance of the data that is explained by the model $f$.

As $R^2$ weighs the deviation at every data point equally, it is not an ideal measure for heteroscedastic data,
like the optical gradient data of \sai.
For this reason, 
we use a generalization of $R^2$,  based on the log-likelihood of the heteroscedastic normal errors, to measure performance in the \sai \ benchmarks.
In particular, the measure is given by
\begin{equation}
    R^2_s 
    = 1 - \frac{\sum_i (f(x_i) - y_i)^2 / \sigma_i^2}
    {\sum_i (\mu(\bold y) - y_i)^2 / \sigma_i^2},
\end{equation}
where $\sigma_i$ is the standard deviation of the $i^{\text{th}}$ error.
For SARA, the $\sigma_i$ are the product of the comprehensive uncertainty quantification of the experimental process.
Clearly, $R^2_s$ reduces to $R^2$ if the noise variances are all equal.
If they are not, 
$R^2_s$ is a better measure of misfit, as it weighs the residuals of more certain data points stronger than those with greater uncertainty. 
Notably, similar pseudo $R^2$ scores based on log-likelihoods are used throughout statistics and applied fields \cite{hu2006pseudo, smith2013comparison, hemmert2018log}.

%*****************************************************************************************
\section{Acknowledgments}\label{sec:ack}
The authors acknowledge the Air Force Office of Scientific Research for support under award FA9550-18-1-0136. This work is based upon research conducted at the Materials Solutions Network at CHESS (MSN-C) which is supported by the Air Force Research Laboratory under award FA8650-19-2-5220,
and the National Science Foundation Expeditions 
under award CCF-1522054. This work was also performed in part at the Cornell NanoScale Facility, a member of the National Nanotechnology Coordinated Infrastructure (NNCI), which is supported by the National Science Foundation (Grant NNCI-2025233). MA acknowledges support from the Swiss National Science Foundation (project P4P4P2-180669). This research was conducted with support from the Cornell University Center for Advanced Computing.

\section{Author contributions}
RBvD, CPG, MOT, and JMG conceived and supervised the research. SA and MA developed and implemented the SARA algorithms, and contributed equally to this work. MA and SA took the lead in writing the manuscript. DRS fabricated the \ce{Bi2O3} thin film samples and collected and analyzed the optical microscopy and reflectance data. ABC performed the lg-LSA experiments. DG and JMG processed the XRD data, and DRS and MCC helped analyze the results. All authors provided critical feedback and helped shape the research, analysis, and manuscript.

%*****************************************************************************************

%\bibliography{references}

\end{document}